Title: A comparison of the Alkire-Foster method and a Markov random field approach in the analysis of multidimensional poverty

Joseph Lam

Great Ormond Street Institute of Child Health

Email: joseph.lam.18@ucl.ac.uk

**Introduction**

Multidimensional measures and indices are increasingly relevant for large-scale studies of social welfare and policy intervention. Rather than focusing on a single metric such as income, researchers and policymakers now recognize that deprivation can manifest across several dimensions—such as health, education, living standards, and social participation—that together define an individual's level of well-being or capability. With the increasing availability of population administrative data, there is scope for evaluating poverty measurement methods that can integrate multiple indicators and effectively classify individuals (or households) as impoverished or not.

Two approaches have emerged as widely discussed yet fundamentally differ in their theoretical foundations and implementation. Firstly, the Alkire-Foster (AF) method, which originated in the capabilities-based framework advocated by Amartya Sen and further developed by Sabina Alkire and James Foster [1,2]. The AF method is designed explicitly for policy-relevant measures of multidimensional poverty. It provides a clear set of deprivation cutoffs for each indicator and a final threshold (e.g., how many indicators must be below their cutoffs) to classify an individual as poor. Rooted in welfare economics and Sen's capability approach, the AF method embodies normative choices: which indicators matter, how they are weighted, and the threshold of $k$ deprivations at which a person is deemed poor. The method's simplicity and transparency make it attractive for policy and communication. For example, a government might define four indicators of housing quality and set a threshold of at least two failures to classify a household as precariously housed. AF handles multiple indicators by summing deprivations into a single index (e.g., the Multidimensional Poverty Index, or MPI [3]). It is relatively agnostic about correlations between indicators, focusing instead on counting how many cutoffs are missed.

Secondly, a Markov Random Field (MRF) approach is grounded in probabilistic graphical modelling and inference techniques commonly used in computational statistics and machine learning [4]. MRFs are general-purpose statistical models for representing and inferring joint probability distributions over interconnected variables. When applied to multidimensional poverty, the MRF approach models not only whether an individual is deprived on each indicator, but also how indicators correlate with one another and with an underlying (latent) poverty status ($Z$). By modelling pairwise or clique-level interactions, MRFs provide a richer representation of dependency structures. In principle, they can capture that "lacking potable water often co-occurs with lacking sanitation services," or that "in households with low income, the likelihood of food insecurity is higher." Instead of purely counting deprivations, MRFs estimate how strongly each deprivation relates to (or is conditionally dependent on) other deprivations and the latent poverty indicator. This explicit modelling of correlation structures means that an MRF can highlight which dimensions tend to cluster.

**Promises and Challenges**

With well-defined cutoffs, AF is straightforward to implement and communicate. Policymakers value its transparency: "45% of individuals are considered multidimensionally poor by this measure,". However, AF does not explicitly assess correlation across indicators and thus may produce either overly inclusive or exclusive classifications if the chosen threshold is not well-calibrated. AF has a relatively good scalability to high dimensional data by simply adding more indicators. The key challenge remains on identifying a threshold *k* that is interpretable and has a strong policy relevance.

MRFs require a more elaborate fitting procedure (estimating factors or potentials), typically done in specialized statistical software (e.g., `pgmpy` in Python [5]). MRFs can reveal the extent of which variables correlate with each another. This might prove particularly useful in large administrative data sets containing rich but interdependent indicators. Computation complexity for MRFs increase tremendously as more factors are included (curse of dimensionality) [6]. Regularisation or other constraints on network structure might be required to avoid over-fitting.

Given these differences, the opportunity to test both methods in a simulation context, for example within large populations where administrative data might eventually be used, may offer valuable insights. The aim of this simulation study is to compare the accuracy and agreement between AF and MRFs in classifying multidimensional poverty.

**Methods**

**Data-Generating Mechanism**

**Population Size and Binary Indicators**
We generate a synthetic population of N=50,000 individuals, each having a latent poverty status $Z \in \{0,1\}$. A fraction $p\_z$ = 0.3 are designated "poor" by construction. We then produce *D*=10 binary indicators $X_1, X_2, ..., X_{10}$, mimicking typical setup for AF method with established individual thresholds for each included factor. We assumed each factor had equal weights. For the poor group (*Z*=1), each indicator has a base deprivation probability (e.g., 0.7), and for the non-poor group (*Z*=0), each indicator has a lower probability (e.g., 0.2). This mimics the assumption that poor individuals are more likely to be deprived across multiple dimensions.

**Introducing Correlation**
Real-world indicators of MPI often show inter-dependence (e.g., lacking water often correlates with lacking sanitation). We incorporate correlation by randomly forcing adjacent indicators to match each other at a relatively small probability (e.g., 10%). This is a simplified representation of correlated deprivations, ensuring that data reflect a realistic pattern of co-occurrence.

**Train/Test Split**
After generating the full dataset, we split it into a 75% training set and 25% test set. This approach allows the MRF method to estimate its potentials or factor tables from training data, then be evaluated on unseen test data.

**Estimand**

**Alkire-Foster Classification**

We assume binary indicators and a predefined threshold-based rule. In this simulation, we define the threshold = 4: If at least 4 out of 10 are flagged, they are considered multidimensionally poor. We do not incorporate any weighting differences among indicators in the present simulation, though that is common in real AF applications.

**Markov Random Field Estimation and Inference**

We adopt a pairwise MRF with nodes for the latent poverty variable *Z* plus each indicator $X_i$. We add edges:

1. (*Z*, $X_i$) for all *i* to capture direct dependence of indicators on the individual's poverty status, and
2. ($X_i$, $X_{i+1}$) for adjacent indicators, creating a chain structure.

This structure is a simplification—one could fully connect all indicators, but that might be computationally more demanding and cause overfitting. Even so, the chain edges help represent how certain indicators cluster or affect one another.

We estimate factor potentials directly from joint counts in the training data. In large-scale administrative data or simulations with tens of thousands of records, zero counts become less frequent. We omit regularisation for clarity, and let the empirical joint frequencies define the MRF potentials. This decision mimics a setting where the training sample is large enough that minimal smoothing is needed.

**Belief Propagation and Thresholding**

For inference, we apply belief propagation to compute $P(Z=1|X_1,...,X_{10})$. If this posterior probability is at least θ, we classify the individual as poor. We chose θ=0.5 with an assumption of equal prior probability of being poor and non-poor, and that false positive and false negative are equally costly. A higher threshold typically reduces false positives while likely increasing false negatives—this trade-off is part of what we compare to the AF approach.

**Methods of analysis**

To assess the robustness of our findings, we repeat the entire data generation, MRF training, and testing procedure 50 times, each with a different random seed. This approach provides a distribution of outcomes (e.g., accuracy, false positive rate, false negative rate) rather than a single run, giving us a mean performance metrics (average accuracy across the 50 runs), and confidence intervals on how stable each method performs to random variations in identifying true poverty status. We then assess the agreement of the classification, and inspect the records where the two methods do not agree.

**Results**

In our testing set, AF method had an estimated mean accuracy of 89.5% (95% Confidence Interval 89.4%-89.6%); MRFs method had an estimated mean accuracy of 75.4% (95% Confidence Interval 75.3%-75.5%). Table 1 described the simulation results from each run. AF method is characterised by its high negative predictive value with very few false-negatives; MRFs method is characterised by its high positive predictive value with very few false-positives (Figure 1).

| seed | acc_mrf | tn_mrf | fp_mrf | fn_mrf | tp_mrf | acc_af | tn_af | fp_af | fn_af | tp_af |
|---|---|---|---|---|---|---|---|---|---|---|
| 0 | 0.75648 | 8815 | 0 | 3044 | 641 | 0.89632 | 7595 | 1220 | 76 | 3609 |
| 1 | 0.7584 | 8805 | 1 | 3019 | 675 | 0.89448 | 7551 | 1255 | 64 | 3630 |
| 2 | 0.75568 | 8796 | 0 | 3054 | 650 | 0.89736 | 7580 | 1216 | 67 | 3637 |
| 3 | 0.74816 | 8728 | 0 | 3148 | 624 | 0.89184 | 7450 | 1278 | 74 | 3698 |
| 4 | 0.7516 | 8750 | 1 | 3104 | 645 | 0.90056 | 7592 | 1159 | 84 | 3665 |
| 5 | 0.7512 | 8764 | 0 | 3110 | 626 | 0.89576 | 7524 | 1240 | 63 | 3673 |
| 6 | 0.75576 | 8793 | 0 | 3053 | 654 | 0.89696 | 7591 | 1202 | 86 | 3621 |
| 7 | 0.74368 | 8637 | 0 | 3204 | 659 | 0.89904 | 7462 | 1175 | 87 | 3776 |
| 8 | 0.75736 | 8789 | 0 | 3033 | 678 | 0.8948 | 7549 | 1240 | 75 | 3636 |
| 9 | 0.74984 | 8682 | 0 | 3127 | 691 | 0.894 | 7447 | 1235 | 90 | 3728 |
| 10 | 0.75272 | 8802 | 0 | 3091 | 607 | 0.89384 | 7565 | 1237 | 90 | 3608 |
| 11 | 0.7512 | 8734 | 1 | 3109 | 656 | 0.89768 | 7535 | 1200 | 79 | 3686 |
| 12 | 0.75432 | 8720 | 1 | 3070 | 709 | 0.89288 | 7451 | 1270 | 69 | 3710 |
| 13 | 0.75304 | 8723 | 1 | 3086 | 690 | 0.89512 | 7479 | 1245 | 66 | 3710 |
| 14 | 0.75632 | 8743 | 2 | 3044 | 711 | 0.89408 | 7497 | 1248 | 76 | 3679 |
| 15 | 0.754 | 8794 | 0 | 3075 | 631 | 0.89632 | 7572 | 1222 | 74 | 3632 |
| 16 | 0.7544 | 8763 | 1 | 3069 | 667 | 0.89768 | 7562 | 1202 | 77 | 3659 |
| 17 | 0.7576 | 8837 | 0 | 3030 | 633 | 0.88896 | 7525 | 1312 | 76 | 3587 |
| 18 | 0.75648 | 8790 | 1 | 3043 | 666 | 0.89192 | 7514 | 1277 | 74 | 3635 |
| 19 | 0.75008 | 8706 | 0 | 3124 | 670 | 0.89976 | 7514 | 1192 | 61 | 3733 |
| 20 | 0.75376 | 8760 | 0 | 3078 | 662 | 0.89448 | 7514 | 1246 | 73 | 3667 |
| 21 | 0.75312 | 8739 | 0 | 3086 | 675 | 0.8968 | 7528 | 1211 | 79 | 3682 |
| 22 | 0.75064 | 8693 | 2 | 3115 | 690 | 0.8948 | 7443 | 1252 | 63 | 3742 |
| 23 | 0.75368 | 8792 | 0 | 3079 | 629 | 0.89464 | 7533 | 1259 | 58 | 3650 |
| 24 | 0.75504 | 8749 | 0 | 3062 | 689 | 0.8972 | 7541 | 1208 | 77 | 3674 |
| 25 | 0.75928 | 8813 | 0 | 3009 | 678 | 0.89592 | 7589 | 1224 | 77 | 3610 |
| 26 | 0.748 | 8700 | 0 | 3150 | 650 | 0.89496 | 7456 | 1244 | 69 | 3731 |
| 27 | 0.75272 | 8745 | 0 | 3091 | 664 | 0.89024 | 7440 | 1305 | 67 | 3688 |
| 28 | 0.75128 | 8716 | 0 | 3109 | 675 | 0.89928 | 7529 | 1187 | 72 | 3712 |
| 29 | 0.75488 | 8732 | 0 | 3064 | 704 | 0.8944 | 7468 | 1264 | 56 | 3712 |
| 30 | 0.75208 | 8705 | 0 | 3099 | 696 | 0.89904 | 7522 | 1183 | 79 | 3716 |
| 31 | 0.75056 | 8725 | 0 | 3118 | 657 | 0.89256 | 7462 | 1263 | 80 | 3695 |
| 32 | 0.75704 | 8789 | 0 | 3037 | 674 | 0.89304 | 7510 | 1279 | 58 | 3653 |
| 33 | 0.75368 | 8788 | 0 | 3079 | 633 | 0.89576 | 7549 | 1239 | 64 | 3648 |
| 34 | 0.75424 | 8809 | 0 | 3072 | 619 | 0.89008 | 7506 | 1303 | 71 | 3620 |
| 35 | 0.7596 | 8835 | 1 | 3004 | 660 | 0.8948 | 7585 | 1251 | 64 | 3600 |
| 36 | 0.75328 | 8794 | 1 | 3083 | 622 | 0.89696 | 7580 | 1215 | 73 | 3632 |
| 37 | 0.7536 | 8787 | 0 | 3080 | 633 | 0.89568 | 7557 | 1230 | 74 | 3639 |
| 38 | 0.75488 | 8750 | 0 | 3064 | 686 | 0.89832 | 7544 | 1206 | 65 | 3685 |
| 39 | 0.75072 | 8705 | 1 | 3115 | 679 | 0.89536 | 7474 | 1232 | 76 | 3718 |
| 40 | 0.7468 | 8648 | 0 | 3165 | 687 | 0.89472 | 7409 | 1239 | 77 | 3775 |
| 41 | 0.75456 | 8730 | 0 | 3068 | 702 | 0.89264 | 7467 | 1263 | 79 | 3691 |
| 42 | 0.75616 | 8820 | 0 | 3048 | 632 | 0.89736 | 7599 | 1221 | 62 | 3618 |
| 43 | 0.75696 | 8741 | 0 | 3038 | 721 | 0.89704 | 7519 | 1222 | 65 | 3694 |
| 44 | 0.74896 | 8681 | 0 | 3138 | 681 | 0.89328 | 7429 | 1252 | 82 | 3737 |
| 45 | 0.75608 | 8766 | 0 | 3049 | 685 | 0.89552 | 7522 | 1244 | 62 | 3672 |
| 46 | 0.76048 | 8859 | 0 | 2994 | 647 | 0.8924 | 7570 | 1289 | 56 | 3585 |

| 47 | 0.75256 | 8764 | 2 | 3091 | 643 | 0.8936 | 7497 | 1269 | 61 | 3673 |
| 48 | 0.75232 | 8710 | 1 | 3095 | 694 | 0.8972 | 7488 | 1223 | 62 | 3727 |
| 49 | 0.76024 | 8809 | 1 | 2996 | 694 | 0.89192 | 7536 | 1274 | 77 | 3613 |

Table 1. Simulation results for all runs, confusion matrix for both AF and MRFs methods. Acc = Accuracy; tn = True Negative; tp = True Positive; fn = False Negative; tn = True Negative; af = Alkire-Foster Method; MRF = Markov Random Field method.

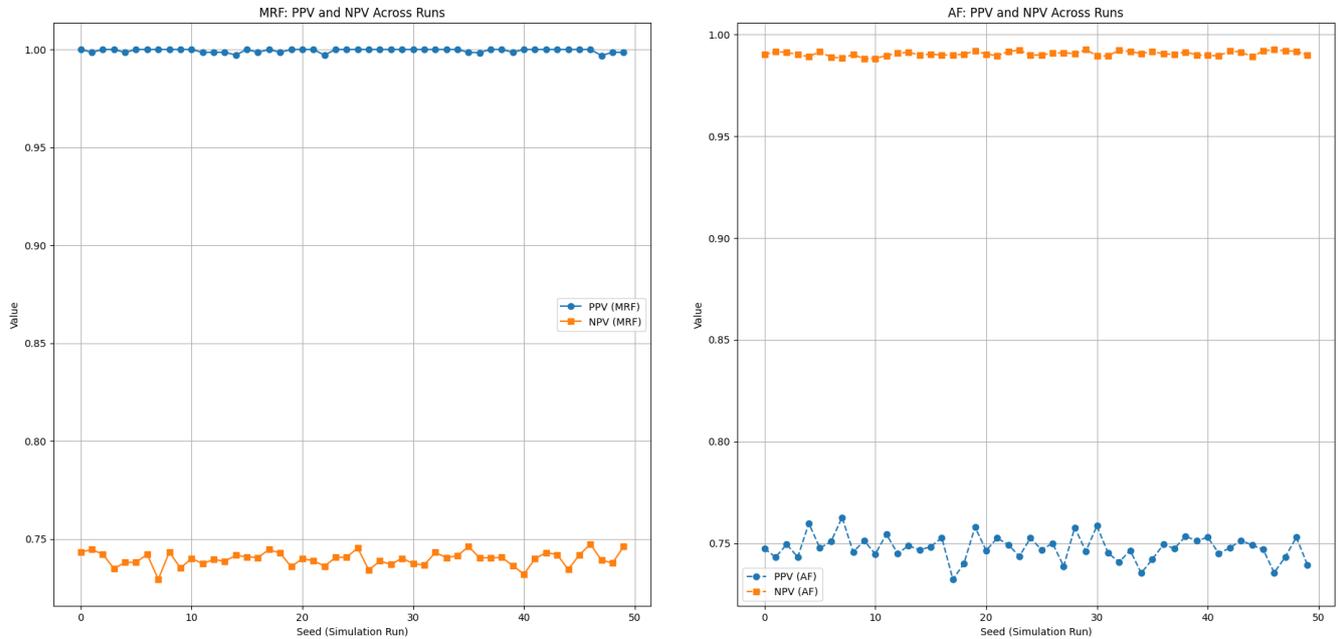

Figure 1. Comparison of Positive Predictive Value (PPV) and Negative Predictive Value (NPV) of simulation runs for MRF (left) and AF (right) methods.

The overall agreement between the two methods is 65%. We further inspected the patterns of disagreement. We included 10 examples when the two methods disagree with each other (Table 2). All disagreements occur when AF = 1, and MRF = 0, given the trade-off described above. MRF tends to class records as "non-poor" when the individual has below-threshold experience at multiple indicators (>4), where AF applies its deterministic threshold (such as IDs 3, 8, 9); AF class the records around the threshold deterministically, whilst MRF can be more flexible in its prediction (such as IDs 1, 2, 5).

| ID | Z | X1 | X2 | X3 | X4 | X5 | X6 | X7 | X8 | X9 | X10 | mrf_pred | af_pred |
|---|---|---|---|---|---|---|---|---|---|---|---|---|---|
| 1 | 0 | 1 | 1 | 0 | 0 | 0 | 0 | 1 | 0 | 1 | 0 | 0 | 1 |
| 2 | 0 | 0 | 1 | 1 | 0 | 1 | 0 | 0 | 0 | 1 | 0 | 0 | 1 |
| 3 | 1 | 1 | 1 | 1 | 1 | 1 | 0 | 1 | 1 | 1 | 0 | 0 | 1 |
| 4 | 0 | 0 | 1 | 1 | 1 | 1 | 1 | 0 | 0 | 0 | 0 | 0 | 1 |
| 5 | 0 | 0 | 1 | 0 | 1 | 1 | 0 | 0 | 1 | 0 | 0 | 0 | 1 |
| 6 | 1 | 0 | 1 | 1 | 1 | 0 | 1 | 1 | 1 | 1 | 1 | 0 | 1 |
| 7 | 0 | 0 | 0 | 1 | 1 | 1 | 0 | 1 | 0 | 0 | 0 | 0 | 1 |
| 8 | 1 | 1 | 1 | 1 | 1 | 0 | 1 | 0 | 1 | 1 | 1 | 0 | 1 |
| 9 | 1 | 1 | 1 | 0 | 1 | 1 | 1 | 1 | 0 | 1 | 1 | 0 | 1 |
| 10 | 1 | 1 | 0 | 1 | 0 | 1 | 0 | 1 | 1 | 1 | 1 | 0 | 1 |

Table 2. Comparison of disagreement patterns. Z = ground-truth (poverty), mrf_pred = MRF prediction, af_pred = AF prediction.

**Discussion**

In this simulation-based comparison, we aim to shed light on how the Alkire-Foster method versus a Markov Random Field approach classify poverty status under large population data scenarios with multiple binary indicators. MRF and AF produced comparable and stable estimation of multidimensional poor status.

By design, the AF method's threshold can lead to different trade-offs. A threshold of 4 (e.g., needing only 4 out of 10 indicators) may cause an inflated false positive rate—many individuals who are not truly poor still meet the partial deprivation count. A high threshold leads to fewer false positives but risks overlooking those with moderate yet critical deprivations (hence, higher false negatives). The method is entirely transparent: one can easily tweak the threshold to match a policy objective (e.g., capturing as many truly poor as possible vs. minimizing over-counting). The MRF's classification depends on the posterior $P(Z=1)$. A threshold of 0.5 might yield fewer false positives if the joint modelling systematically sets the posterior below 0.5 for most ambiguous cases. In practice, we might see the MRF be more conservative if the learned potentials strongly favour non-poor states unless multiple strongly correlated indicators push the posterior above 0.5. MRF can be tuned by adjusting the posterior cutoff but is also influenced by correlations among indicators in the training data, possibly revealing a dimension that correlates strongly with $Z$ and overshadowing others. The acceptability of false positives or negatives depends on policy goals, as well as the context or population the method is applied. If the concern is ensuring no truly poor individual is excluded, a lower threshold for AF or a lower posterior cutoff for MRF is warranted. If the concern is focusing resources on the most severely deprived, a higher threshold or posterior cutoff is beneficial.

Both methods, as set up here, rely on binary indicators. In practice, many relevant deprivation measures (e.g., income, BMI, years of education) are continuous or ordinal. By converting them to "above/below cutoff," we lose information. This can be convenient—some policy thresholds (like "below the national poverty line" or "less than 5 years of education") are conceptually straightforward. However, from a statistical perspective, binarizing might artificially reduce variance or misclassify borderline individuals. The MRF approach could, in theory, be extended to handle ordinal/continuous variables, either directly (via other distributions) or by using more granular categories. Similarly, the AF method can be adapted if some thresholds are chosen at multiple cutoff levels. Yet, simplicity and policy clarity often motivate binary approaches, accepting some potential distortion.

In comparison, MRFs is more computationally intensive but also more flexible, as it offers deeper insight into how deprivation indicators correlate. In large administrative datasets, it might be particularly informative to see which clusters of indicators strongly predict an individual's latent poverty status. In large population data contexts (like administrative records linking social welfare usage, health insurance claims, and housing quality data), the MRF approach can highlight complex interplay: for instance, individuals lacking stable housing plus high rates of emergency department visits might shape the posterior probability of "latent poverty" more strongly than any single indicator alone. The AF method would still capture the presence of multiple deprivations, but it would not systematically quantify how these indicators reinforce each other's likelihood. Nonetheless, the interpretability hurdle for MRFs is higher, and the need for specialized software or parameter tuning (e.g. smoothing if data are sparse) should not be overlooked.

Both methods scale to large data. AF scales linearly with the number of individuals, while MRF inference can grow in complexity, though pairwise MRFs with efficient inference algorithms (like belief propagation) remain tractable for large data if carefully implemented. In real administrative settings, the complexities of data cleaning, missing values, or changes in definitions over time add further challenges. The emphasis on "no smoothing" in our simulation underscores the assumption of abundant data, but real applications may require Bayesian priors or regularization in sparse contexts.

**Conclusion**

In summary, both the Alkire-Foster method and the Markov Random Field approach address the challenge of identifying multidimensional poverty within large population datasets, but with distinct theoretical foundations and practical trade-offs.

By applying a simulation framework with repeated runs and large synthetic populations, we gain insight into the classification behaviour of each method under controlled conditions. This work lays the foundation for implementation of the two methods using real-world data. This exercise highlights the differences in false positive/negative tendencies, the consequences of binarizing data, and the practical considerations of each method's implementation.

In real-world usage, a researcher or policymaker might choose AF for its simplicity and policy resonance, or MRF for its richer structure and potential to uncover interactions among deprivations. Ultimately, the best choice may depend on data availability, technical capacity, and whether the goal is broad poverty measurement or a more detailed exploration of which combinations of deprivations drive capability shortfalls.

**Reference**

1. Alkire S, Roche JM, Ballon P, Foster J, Santos ME, Seth S. Multidimensional Poverty Measurement and Analysis. Oxford University Press; 2015. 369 p.

2. Nussbaum MC, Sen A. The Quality of Life. Oxford University Press; 1993. 466 p.

3. Alkire S, Santos ME. Measuring Acute Poverty in the Developing World: Robustness and Scope of the Multidimensional Poverty Index. World Development. 2014 Jul 1;59:251–74.

4. Metzler D, Croft WB. A Markov random field model for term dependencies. In: Proceedings of the 28th annual international ACM SIGIR conference on Research and development in information retrieval [Internet]. New York, NY, USA: Association for Computing Machinery; 2005 [cited 2025 Mar 4]. p. 472–9. [SIGIR '05]. Available from: https://dl.acm.org/doi/10.1145/1076034.1076115

5. Ankan A, Panda A. pgmpy: Probabilistic Graphical Models using Python. In Austin, Texas; 2015 [cited 2025 Mar 4]. p. 6–11. Available from: https://doi.curvenote.com/10.25080/Majora-7b98e3ed-001

6. Koppen M. The curse of dimensionality.